\documentclass[12pt,twocolumn]{article} 
\usepackage{amsmath}
\usepackage{amssymb} 
\usepackage[dvips]{graphicx} 
\begin{document}

\title{On the Solutions of the Lorentz-Dirac Equation}
\author{D. Vogt\thanks{e-mail: danielvt@ifi.unicamp.br}\\
Instituto de F\'{\i}sica Gleb Wataghin, Universidade
 Estadual de Campinas\\
13083-970 Campinas, S.P., Brazil
\and
P. S. Letelier\thanks{e-mail: letelier@ime.unicamp.br}\\
Departamento de Matem\'{a}tica Aplicada-IMECC,
 Universidade Estadual\\
de Campinas 13083-970 Campinas, S.P., Brazil}
\date{}
\maketitle

\noindent
{\bf  Abstract}
\vspace{0.3cm}

\noindent
We discuss the unstable character of the solutions of the
Lorentz-Dirac equation and stress the need of  methods 
like order reduction
to derive a  physically acceptable  equation of motion.
 The discussion is illustrated
with the paradigmatic example of the non-relativistic harmonic
oscillator with radiation reaction. We also illustrate removal of
 the noncasual pre-acceleration 
with the introduction of a small correction in the 
Lorentz-Dirac equation.
\vspace{0.1cm}

\noindent
{\it PACS: }  41.60.-m,  02.60.Cb  \\
\noindent
{\it Keywords:} Lorentz-Dirac equation, radiation reaction, stability,
 harmonic oscillator.

\noindent
\hrulefill \hspace{3cm}

\hspace{1.5cm}

The classical equation of motion of a point charge including
radiation reaction is the Lorentz-Dirac (LD) equation \cite{Dirac},
\begin{equation} \label{eq_Dirac}
\ddot{x}^{\mu }=f^{\mu}+ b\left( \dddot{x}^{\mu }+ 
\frac{\dot{x}^{\mu }}{c^{2}}\ddot{x}^{\nu }\ddot{x}_\nu \right) \mbox{,}  
\end{equation} 
where a dot indicates derivative with respect to the proper time $\tau$;
$f^{\mu}$ is the external 4-force per unit mass and $b=\frac{2e^2}{3mc^3}$.
Although the LD equation is based in  solid physical ground
(Special Relativity 
and Electrodynamics) it has some  unusual mathematical properties that
 put some doubts about the 
suitability of this equation to represent the motion equation of a charged 
particle
with radiation reaction.
Since the LD equation  is a third order ordinary differential equation, 
not only the initial position
and velocity, but also the initial acceleration must be given to specify
a unique solution. Moreover, one has to determine with an infinity
 accuracy the initial value of the
acceleration  that  eliminates the  part of the solutions that 
grow exponentially in time (``runaway'' solution).  

Recently, Spohn \cite{Spohn} found that the physical solutions of the LD equation
lie on a critical manifold in phase space that consists exclusively of
repelling fixed points. On this critical surface the motion is free of
runaway solutions and is governed by an effective second order equation. If
an initial condition slightly off the critical surface is given, 
the solution 
grows exponentially fast. This lack of stability of solutions with 
respect to small
variations of initial conditions means that the LD equation does
 not represent
a mathematically well posed problem \cite{Zauderer}  and in
 consequence its applicability is limited.

Difficulties also arise in the numerical solution of the LD
 equation \cite{Baylis}. Even if
 one knows the correct
initial value of the acceleration, in the usual forward numerical integration
the runaway contributions to the solution grow extremely fast due to numerical errors. 
As an alternative, Aguirregabiria \cite{Aguirre}  proposes a method of order reduction in
which Eq.(\ref{eq_Dirac}) is substituted by a second order equation of motion
with no runaway solutions. The explicit form of this second order equation
cannot be obtained in general, so a series of successive approximations 
$\ddot{x}^{\mu}=\xi_n^{\mu}$, $(n=0,1,\ldots)$ is constructed:
\begin{align}
\ddot{x}^{\mu} &=\xi_0^{\mu}=f^{\mu}, \label{eq_aprox1} \\ 
\ddot{x}^{\mu} &=\xi_1^{\mu}=f^{\mu}+b\left( \frac{\partial f^{\mu}} 
{\partial \tau}+\frac{\partial f^{\mu}}{\partial x^{\nu}} \dot{x}^{\nu}+ 
\frac{\partial f^{\mu}}{\partial \dot{x}^{\nu}} f^{\nu}\right. \nonumber \\
&\left. +\dot{x}^{\mu} \frac{f^{\nu}f_\nu}{c^{2}} \right), \label{eq_aprox2} \\ 
\ddot{x}^{\mu} &=\xi_{n+1}^{\mu}=f^{\mu}+b\left( \frac{\partial \xi_n^{\mu}} 
{\partial \tau}+\frac{\partial \xi_n^{\mu}}{\partial x^{\nu}} \dot{x}^{\nu}+ 
\frac{\partial \xi_n^{\mu}}{\partial \dot{x}^{\nu}} \xi_n^{\nu}\right. \nonumber \\ 
&\left. +\dot{x}^{\mu} \frac{\xi_n^{\nu}\xi_{n \nu}}{c^{2}} \right) \mbox{.} \label{eq_aprox3}
\end{align}
We note that in this approach only the usual initial 
conditions (position and velocity) 
are required and the equations  for a {\it finite} $n$ are stable with respect to variations
of initial conditions.

Rohrlich \cite{Rohrlich}, inspired on Landau \cite{Landau},  argues
that the original LD equation should be replaced by Eq.(\ref{eq_aprox2})
as the \emph{exact} equation of motion of a radiating point particle, since
it is of second order, satisfies
the principle of inertia, and the radiation reaction term vanishes in the
absence of an external force. This
prescription is based on physical arguments, but only the confrontation with
experimental results will establish 
its validity. For proposals of experimental tests, see \cite{Spohn}.

The purpose of this letter is to illustrate the above discussion with
the significant example of the non-relativistic one-dimensional harmonic 
oscillator with radiation reaction.
This problem has an  analytical solution that can be used to construct the
effective second order equation of motion on the above mentioned  critical
 surface. The successive
approximations scheme (\ref{eq_aprox1})--(\ref{eq_aprox3})  and solutions 
from direct numerical
integration of the non-relativistic LD equation  can  be confronted with
 the exact solution.
The study of this example will show us the limitations of the LD equation

 to describe the motion of a
charge particle interacting  with its  own radiation (back reaction). Also
 we shall discuss the 
addition of a small correction term that eliminates the pre-acceleration 
in the solutions to the 
LD equation \cite{Yaghjian}. In particular we discuss, in some detail, 
the case of a constant force 
that acts on the charge particle during a finite time.
 
The one-dimensional non-relativistic motion of an electron
 subjected to a harmonic 
force is governed by the Abraham-Lorentz limit of the LD equation,
\begin{equation} \label{eq_harm}
\frac{d^2x}{dt^2}=-\omega^{2}x+b\frac{d^3x}{dt^3} \mbox{,}
\end{equation}
whose exact solution is given by
\begin{equation} \label{sol_osc_harm} 
x(t)=c_1e^{\alpha_1t}+e^{\alpha_{2r}t} \left( c_2\cos(\alpha_{2i}t)+ 
c_3 \sin(\alpha_{2i}t) \right) \mbox{.} 
\end{equation}
The constants $c_1$, $c_2$ and $c_3$ depend on the initial conditions and 
\begin{gather} 
\alpha_1=\frac{1}{3b} \left( 1+\beta^{1/3}+\beta^{-1/3}
 \right) \mbox{,} \\ 
\alpha_{2r}=\frac{1}{3b} \left( 1-\frac{\gamma^{1/3}}{4}
-\gamma^{-1/3} \right) \mbox{,} \\ 
\alpha_{2i}=\frac{\sqrt{3}}{6b} \left( \beta^{1/3}-
\beta^{-1/3} \right) \mbox{,} \\ 
\beta=1+\frac{27}{2}\omega^2 b^2+3\sqrt{3}\omega b 
\sqrt{1+\frac{27}{4}\omega^2 b^2} \mbox{,} \\ 
\gamma=8+108\omega^2 b^2+12\sqrt{3}\omega b \sqrt{4+27\omega^2 b^2} \mbox{.} 
\end{gather} 

>From Eq.(\ref{sol_osc_harm}) we find  the expressions for the
velocity and acceleration:
\begin{gather} 
\dot{x}(t)=\alpha_1c_1e^{\alpha_1t}+(c_2\alpha_{2r}+c_3\alpha_{2i})
e^{\alpha_{2r}t}\cos(\alpha_{2i}t)\nonumber\\ 
+(c_3\alpha_{2r}-c_2\alpha_{2i})e^{\alpha_{2r}t}\sin(\alpha_{2i}t)
 \mbox{,} \label{eq_veloc} \\ 
\ddot{x}(t)=\alpha_1^2c_1e^{\alpha_1t}+\left[ c_2(\alpha_{2r}^2-
\alpha_{2i}^2)+2c_3\alpha_{2r}\alpha_{2i}\right]\times \nonumber\\ 
 e^{\alpha_{2r}t}\cos(\alpha_{2i}t)+
\left[ c_3(\alpha_{2r}^2-\alpha_{2i}^2) \right. \nonumber\\ 
\left. -2c_2\alpha_{2r}\alpha_{2i}\right] e^{\alpha_{2r}t}
\sin(\alpha_{2i}t) \mbox{.} \label{eq_acel}  
\end{gather} 
In terms of the initial conditions $x(0)=x_0$, $\dot{x}(0)=\dot{x}_0$
 and $\ddot{x}(0)=\ddot{x}_0$, the constants $c_1$, $c_2$ and $c_3$ are given by
\begin{gather} 
c_1=\frac{x_0 \left( \alpha_{2r}^2+\alpha_{2i}^2 \right)+\ddot{x}_0
-2\alpha_{2r}\dot{x}_0}{\left( 
\alpha_1-\alpha_{2r} \right)^2+\alpha_{2i}^2} \mbox{,} \label{eq_cond1} \\ 
c_2=\frac{x_0\alpha_1 \left( \alpha_1-2\alpha_{2r} 
\right)-\ddot{x}_0+2\alpha_{2r}\dot{x}_0}{\left( \alpha_1-\alpha_{2r} 
\right)^2+\alpha_{2i}^2} \mbox{,} \label{eq_cond2} \\ 
c_3=-\left[ x_0\alpha_1 \left( \alpha_1\alpha_{2r}-\alpha_{2r}^2+
\alpha_{2i}^2\right)+\right. \nonumber \\
\left. +\ddot{x}_0 \left( \alpha_1-\alpha_{2r} \right)+\dot{x}_0
 \left( \alpha_{2r}^2-\alpha_1^2-
\alpha_{2i}^2 \right) \right]/ \nonumber \\
\alpha_{2i} \left[ \left( \alpha_1-\alpha_{2r} \right)^2+\alpha_{2i}^2 
\right] \mbox{.} \label{eq_cond3}  
\end{gather} 

Using Eq.(\ref{sol_osc_harm}) and Eq.(\ref{eq_veloc}), Eq.(\ref{eq_acel}) can
be cast  as
\begin{multline} \label{eq_inter}
\ddot{x}(t)=c_1 \left[ (\alpha_1-\alpha_{2r})^2+\alpha_{2i}^2
 \right] e^{\alpha_1t}+2\alpha_{2r}\dot{x}\\
-(\alpha_{2r}^2+\alpha_{2i}^2)x \mbox{.} 
\end{multline}  
Since $\alpha_1>0$, the term $e^{\alpha_1t}$ has a runaway
 character, so we
 must set $c_1=0$ to eliminate the unphysical solutions. 
Then Eq.(\ref{eq_inter}) takes the form
\begin{equation} \label{eq_efetiva} 
\ddot{x}(t)=2\alpha_{2r}\dot{x}-(\alpha_{2r}^2+\alpha_{2i}^2)x \mbox{.} 
\end{equation}
This is the equation of motion of a damped harmonic oscillator and represents
the effective second order equation on the critical surface. In this particular
problem the effective equation can be obtained
exactly.

The scheme of successive approximations applied to Eq.(\ref{eq_harm})
 results in equations similar to Eq.(\ref{eq_efetiva}) \cite{Aguirre}
\begin{equation} \label{eq_reduzida} 
\ddot{x}=-\gamma_n\dot{x}-\omega_n^2x \text{ , }n=0,1,... 
\end{equation} 
where the coefficients satisfy the recurrence relations 
\begin{gather} 
\gamma_0=0 \mbox{,}\nonumber \\ 
\omega_{n+1}^2=\omega_0^2-b\omega_n^2\gamma_n \mbox{,}\\ 
\gamma_{n+1}=b \left( \omega_n^2-\gamma_n^2 \right) \mbox{.} \nonumber 
\end{gather} 
When $b\omega\lessapprox 0.95$ these successive approximations
  will always converge.

\begin{figure}  
\centering 
\fbox{\includegraphics[scale=0.6]{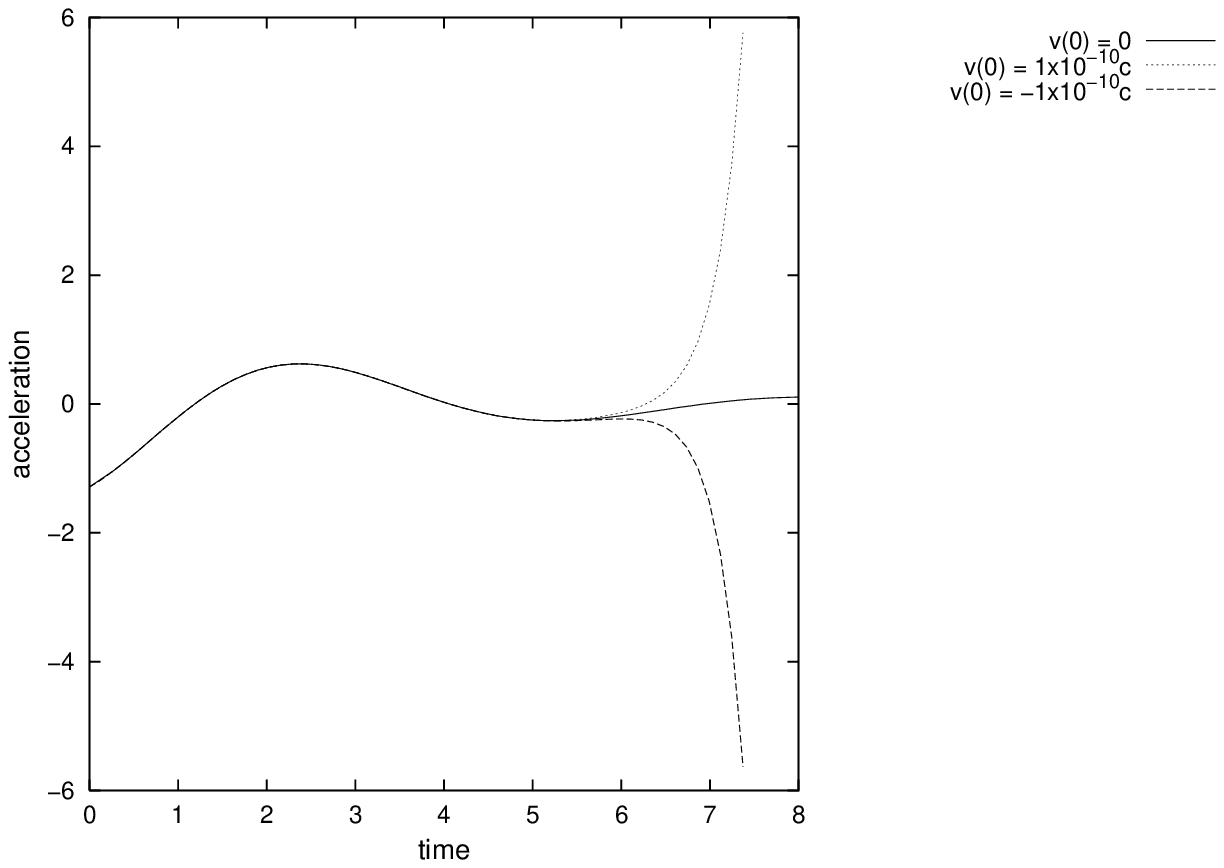}} 
\caption{Acceleration as function of time with $b\omega=2\pi/7$ and initial 
conditions $x(0)=1$, $\dot{x}(0)=0$ (solid curve),
 $\dot{x}(0)=\pm 1 \times 10^{-10}$c (dashed curves) and $\ddot{x}(0)$ 
given by Eq.(\ref{eq_efetiva}). The curves were ploted  using   the analytical
 solutions of Eq.(\ref{eq_harm}) (1 unit of length $=2.818 \times 10^{-15}$ m, 1 unit
 of time $=9.399 \times 10^{-24}$ s).} 
\end{figure} 

Figure 1 shows the curves of acceleration as function of time 
with $x(0)=1$,  $b\omega=2\pi/7$ 
(our units are: 1 unit of length $=2.818 \times 10^{-15}$ m, 1 unit 
of time $=9.399 \times 10^{-24}$ s). The continuous curve was obtained 
from Eq.(\ref{eq_efetiva}) with initial velocity $\dot{x}(0)=0$, the dashed 
curves with the same initial values for position and acceleration, but
 with $\dot{x}(0)=\pm 1 \times 10^{-10}$c and Eq.(\ref{eq_acel})
--(\ref{eq_cond3}). This
 is equivalent to take initial conditions slightly off the critical
 surface. The
 acceleration diverges after a few units of   time, indicating the
 instability of the critical surface. 

\begin{figure}  
\centering 
\fbox{\includegraphics[scale=0.6]{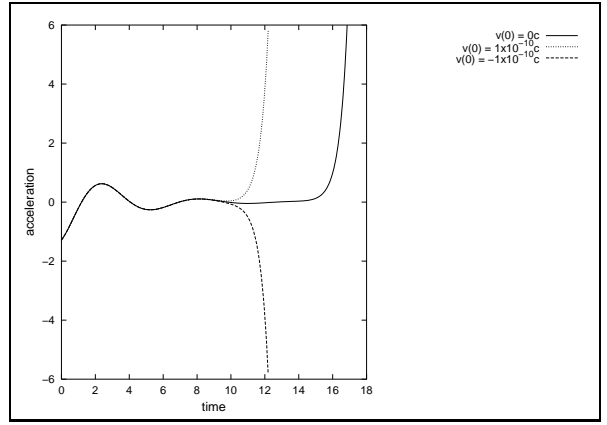}} 
\caption{Acceleration as function of time with $b\omega=2\pi/7$ and 
initial conditions $x(0)=1$ unit        
 of length, $\dot{x}(0)=0$ (solid curve), $\dot{x}(0)=\pm 1 
\times 10^{-10}$c (dashed curves)
 and $\ddot{x}(0)$ given by Eq.(\ref{eq_efetiva}). The curves were
 drawn   numerically integrating 
 Eq.(\ref{eq_harm}) (1 unit of length $=2.818 \times 10^{-15}$ m, 
1 unit of time $=9.399 \times 10^{-24}$ s).} 
\end{figure} 

Figure 2 shows the curves of acceleration as function of time 
with $b\omega=2\pi/7$ and the same 
initial conditions as  in figure 1, but now Eq.(\ref{eq_harm}) was
 solved  numerically
  (fourth-fifth order Runge-Kutta with adaptive step algorithm). Even
 though the value of
 the initial acceleration given by Eq.(\ref{eq_efetiva}) is used with
 16 significant figures, the
 runaway contributions make also the numerical solution useless after
 a few time units.

\begin{figure}  
\centering 
\fbox{\includegraphics[scale=0.6]{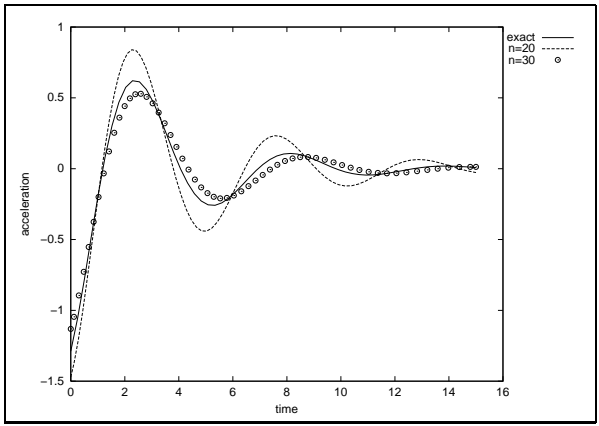}} 
\caption{Acceleration as function of time with $b\omega=2\pi/7$ and initial condition
 $x(0)=1$ unit of length and $\dot{x}(0)=0$. The curves were plot  using  successive
 approximations Eq.(\ref{eq_reduzida}) 
for $n=20$ and $n=30$ (1 unit of length $=2.818 \times 10^{-15}$ m, 1 unit of
 time $=9.399 \times 10^{-24}$ s).} 
\end{figure} 

Finally, in Figure 3 the solutions calculated with the successive
 approximations (\ref{eq_reduzida})
 are compared with the exact solution using initial conditions
 $x(0)=1$ and $\dot{x}(0)=0$. Although
 the convergence is slow, no instability is seen.

This example clarifies the uselessness of Eq.(\ref{eq_Dirac}) 
in practice. Even though, for very special cases, 
 some exact solutions of  the LD equation are known \cite{Plass}
 for actual situations one must rely on
numerical solutions. But, to eliminate the runaway components one 
would have to know the {\it exact} initial conditions and to perform 
the computations with {\it infinite} precision. The knowledge of the 
exact initial conditions imposes a physically 
impossible condition to be achieved. Also, do numerics with infinite
 precision is not possible.
Recently, Chicone \textit{et al} \cite{Chicone} pointed out that 
 high order equations like Eq.(\ref{eq_Dirac})
must be viewed as an intermediate step in the derivation of the physically 
correct, second order equation. 
A possible method to find this equations is  the use of successive 
approximations. However, there is no guarantee 
of convergence.

We would like to comment another unusual feature of the LD equation, 
the phenomenon of pre-acceleration.
When the external force is an explicit function of time, the solution 
of the one-dimensional
non-relativistic version of Eq.(\ref{eq_Dirac}) can be written as \cite{Plass}
\begin{equation} \label{eq_accel_1}
\ddot{x}(t)=\frac{1}{b}\int_t^{\infty}e^{-(t'-t)/b}f(t') \, \mathrm{d}t' \mbox{.}
\end{equation}
It follows from Eq.(\ref{eq_accel_1}) that a nonzero acceleration exists
 before the external
force is applied at $t=0$, thus violating cusality. In a nice monograph
 on the LD equation, Yaghjian \cite{Yaghjian}
carefully rederives the equation of motion for the extended model of the
 electron as a charged
insulating sphere of radius $a$. He shows that the multiplication of
 the electromagnetic self force
by a correction function $\eta(\tau)$ eliminates the pre-acceleration
 from the solution to the
original LD equation without introducing false discontinuities in
 velocity across $\tau=0$ or
spurious delta functions and their derivatives at $\tau=0$. The
 correction function increases
monotonically from zero to one in the time it takes light to travel 
across the electron, and
it approaches zero like $\tau^2$ or faster as $\tau$ approaches
 zero from the positive (right) side. 
This modification is needed to ensure the validity of the Taylor
 expansion that is used in the 
evaluation of the electromagnetic self force about present
 time $(\tau)$ of position, velocity and 
acceleration of each element of charge at retarded time
 during the interval $0<\tau<2a/c$. 
With this correction, Eq.(\ref{eq_accel_1}) is modified to
\begin{equation} \label{eq_sol_mod}
\ddot{x}(t) =\begin{cases} 0 & \text{, $t<0$} \\
-\int_t^{\infty} f(t') \frac{d}{dt'}
\left[ e^{-\frac{1}{b}\int_t^{t'}\frac{d t''}{\eta (t'')}} \right]
\, \mathrm{d}t' & \text{, $t \geq 0$.} \end{cases}
\end{equation}

A simple example helps to clarify the differences 
between Eq.(\ref{eq_accel_1}) and Eq.(\ref{eq_sol_mod}). 
Let $f(t)$ be a constant force that acts during a finite period of time,
\begin{equation}
f(t)= \begin{cases} 0 & \text{, $0<t<t_0$} \\
k & \text{, $t_0<t<t_1$} \\
0 & \text{, $t>t_1$.} \end{cases}
\end{equation}
The solution Eq.(\ref{eq_accel_1}) is given by \cite{Plass}
\begin{equation} \label{eq_sol_acel1}
\ddot{x}(t)= \begin{cases} ke^{t/b}(e^{-t_0/b}-e^{-t_1/b}) &
\text{, $0<t<t_0$} \\
k[1-e^{-(t_1-t)/b}] & \text{, $t_0<t<t_1$} \\
0 & \text{, $t>t_1$.} \end{cases}
\end{equation}
If we assume following form for the correction function $\eta(t)$:
\begin{equation}
\eta(t)=\begin{cases} 0 & \text{, $0<t<t_0$} \\
\frac{(t-t_0)^2}{4a^2} & \text{, $t_0<t<t_0+2a$} \\
1 & \text{, $t>t_0+2a$,} \end{cases} 
\end{equation}
where we suppose that $t_0+2a<t_1$, Eq.(\ref{eq_sol_mod}) can
 be evaluated exactly. We get
\begin{equation} \label{eq_acel_mod}
\ddot{x}(t)= \begin{cases} 0 & \text{, $0<t<t_0$} \\
k \left[ 1-e^{\frac{4a}{b}\left( 1-\frac{a}{t-t_0} \right) -\frac{(t_1-t_0)}
{b}} \right] \text{,} & \\
\text{ $t_0<t<t_0+2a$} & \\
k[ 1-e^{-(t_1-t)/b}] \text{,} & \\
\text{ $t_0+2a<t<t_1$} & \\
0 & \text{, $t>t_1$.} \end{cases}
\end{equation}

Eq.(\ref{eq_acel_mod}) shows there is no pre-acceleration
 in the interval $0<t<t_0$ before the
external force begins to act. In the limit $t \rightarrow t_0^+$, 
solution (\ref{eq_acel_mod}) 
reduces to $\ddot{x}(t_0)=k$, i. e., the acceleration equals
 the external force per unit mass
when the external force is first applied, a result that
 Yaghjian shows is valid in general.
Figure 4 displays the curves of accelerations calculated
 with Eq.(\ref{eq_sol_acel1}) and 
Eq.(\ref{eq_acel_mod}) for $t_0=2\text{, }t_1=5\text{, }a=
1\text{, and }k=0.1$. When $\eta(t)$
reduces to 1, after $t=t_0+2a$, both solutions agree.
\begin{figure}  
\centering 
\fbox{\includegraphics[scale=0.6]{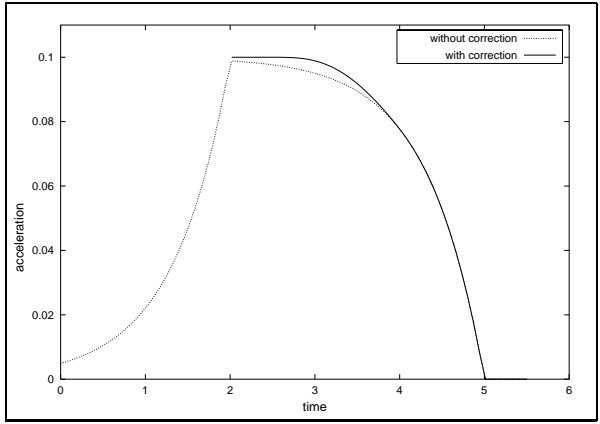}} 
\caption{Accelerations calculated with Eq.(\ref{eq_sol_acel1}) 
and Eq.(\ref{eq_acel_mod})
for $t_0=2\text{, }t_1=5\text{, }a=1\text{, }k=0.1$, (1 unit of
 length $=2.818 \times 10^{-15}$ m, 1 unit
 of time $=9.399 \times 10^{-24}$ s). Note the elimination of
 pre-acceleration when
the correction function is included in the Abraham-Lorentz 
equation of motion.} 
\end{figure}

To summarize, the LD equation suffers from two major problems:
 pre-acceleration and runaway solutions. 
To have a physically acceptable equation of motion two

 approaches have been proposed. In the first, we 
introduce a correction function multiplying the electromagnetic
 self-force term in the original LD 
equation to eliminate the pre-acceleration \cite{Yaghjian} and
 impose a boundary condition to eliminate 
the runaway solution. And in the second by an appropriate
 procedure we substitute the third order 
differential LD equation by an effective second order
 differential equation \cite{Landau}.
Rohrlich \cite{Rohrlich} advocate to take this effective
 second order equation as the right equation 
for the charged particle motion.

\vspace{0.3cm}
\noindent
{\it Acknowledgments.} D.V. thanks \textsc{capes} 
  and P.S.L. thanks  \textsc{fapesp} and \textsc{cnpq} for
 financial support.

\end{document}